\documentstyle[twoside,11pt]{article}
\normalsize
\def\greaterthansquiggle{\raise.3ex\hbox{$>$\kern-.75em\lower1ex\hbox{$\sim$}}}
\def\lessthansquiggle{\raise.3ex\hbox{$<$\kern-.75em\lower1ex\hbox{$\sim$}}}
\newcommand{\bdi}{\begin{displaymath}}
\newcommand{\edi}{\end{displaymath}}
\newcommand{\bfi}{\begin{figure}}
\newcommand{\efi}{\end{figure}}

\newcommand{\beq}{\begin{equation}}
\newcommand{\eeq}{\end{equation}}

\newcommand{\gaM}{\gamma^{\mu}}

\newcommand{\beqa}{\begin{eqnarray}}
\newcommand{\eeqa}{\end{eqnarray}}

\newcommand{\ra}{\rightarrow}

\def\au{{\setbox0=\hbox{\lower1.36775ex%
\hbox{''}\kern-.05em}\dp0=.36775ex\hskip0pt\box0}}
\def\ao{{}\kern-.10em\hbox{``}}

\newcommand{\dsla}{\partial\hspace{-6pt} /  }

\newcommand{\ddsla}{\partial\hspace{-4.6pt} /  }

\newcommand{\AAsla}{A\hspace{-5pt}  /  }

\begin{document}
\bibliographystyle{plain}

\begin{titlepage}
\begin{flushright} 
PM 96/01
\end{flushright}

\bigskip

\bigskip

\begin{center}

{\Large \bf  The boson-boson bound state in the massive Schwinger model}\\

\bigskip

\bigskip

{\bf Christoph Adam} \\
Laboratoire de Physique Math\'ematique, \\
Universit\'e de Montpellier II, Place Eug\`ene Bataillon, \\
34095 - Montpellier Cedex 05, France.$^*)$ \\

\bigskip

\bigskip

\bigskip

\bigskip

{\bf Abstract} \\

\end{center}

\bigskip

\medskip

We use (fermion) mass perturbation theory for the massive Schwinger model
to compute the boson-boson bound state mass in lowest order.
For small fermion mass the lowest possible Fock state turns out to give the 
main contribution and leads to a second order result for the bound state mass.

\vfill

$^*)${\footnotesize permanent address: Institut f\"ur theoretische Physik,
Universit\"at Wien \\
Boltzmanngasse 5, 1090 Wien, Austria \\
email address: adam@pap.univie.ac.at}
\end{titlepage}

\section{Introduction}

The massless Schwinger model - which is two-dimensional QED with one massless 
fermion -  is wellknown to be exactly soluble (\cite{Sc1} - \cite{SW1}, 
\cite{Adam} - \cite{ABH}).
The spectrum of the model consists of one free, massive boson with Schwinger 
mass $\mu_0^2 =\frac{e^2}{\pi}$, which may be interpreted as a 
fermion-antifermion bound state (\cite{IP1}, \cite{CKS}). 
Besides, instantons and a nontrivial vacuum 
structure ($\theta$-vacuum) are present, and a fermion condensate is formed 
(\cite{DSEQ} - \cite{Sm2}).

The massive Schwinger model (with one massive fermion) is no longer exactly 
soluble, and the physical particle (the Schwinger boson) is no longer free
(\cite{KS1} - \cite{Fry}). 
However, the nontrivial features of the massless model (instantons, 
$\theta$-vacuum, fermion condensate) persist to be present in the massive case
(\cite{CJS} - \cite{SMASS}). 

The known exact solution of the massless model can be used for a mass 
perturbation expansion of the massive theory that preserves all the nontrivial
 features of the model (\cite{FS1}). 
For the fermion condensate and Schwinger mass this
was done in \cite{MSSM}, \cite{SMASS}.

Here we focus on the boson-boson interaction. It is mediated by an attractive 
force and therefore a boson-boson bound state is formed. We compute the mass of
this bound state within mass perturbation theory in leading order.

\section{Exact $n$-point functions of the massless model}

The vacuum functional (and Green functions) of the massive Schwinger model may
be inferred from $n$--point functions of the massless Schwinger model via an
expansion in the fermion mass. Indeed, we may write for the Euclidean vacuum
functional ($k\ldots$ instanton number)
\beq
Z(m,\theta)=\sum_{k=-\infty}^{\infty}e^{ik\theta}Z_k (m)
\eeq
where
\bdi
Z_k (m)=N\int D\bar\Psi D\Psi DA^\mu_k e^{\int dx\Bigl[ \bar\Psi(i\ddsla
-e\AAsla_k +m)\Psi -\frac{1}{4}F_{\mu\nu}F^{\mu\nu}\Bigr] }
\edi
\bdi
=N\int D\bar\Psi D\Psi D\beta_k \sum_{n=0}^\infty \frac{m^n}{n!}\prod_{i=1}^n
\int dx_i \bar\Psi (x_i)\Psi (x_i)\cdot
\edi
\beq
\cdot \exp\{\int dx\Bigl[ \bar\Psi (i\dsla -\epsilon_{\mu\nu}\gaM\partial^\nu \beta_k
)\Psi +\frac{1}{2e^2}\beta_k {\Box}^2 \beta_k \Bigr] \}
\eeq
($eA_\mu =\epsilon_{\mu\nu}\partial^\nu \beta$ corresponding to Lorentz gauge).
Therefore, the perturbative computation of $Z(m,\theta)$ is traced back to the
computation of scalar VEVs ( $\langle\prod_i S(x_i )\rangle_0 $, $S(x)\equiv
\bar\Psi (x)\Psi (x)$ ) for the massless Schwinger model and some space time
integrations. It is useful to rewrite the scalar densities in terms of chiral
ones, $S(x)=S_+ (x)+S_- (x)$, $S_\pm \equiv \bar\Psi P_\pm \Psi$, because in
this case
only a definite instanton sector $k=n_+ -n_-$ contributes to the VEV
$\langle\prod_{i=1}^{n_+} S_+ (x_i)\prod_{j=1}^{n_-} S_- (x_j) \, \hat O 
\rangle_0$, where $\hat O$ is a chirality neutral operator (e.g. a product of
vector currents). A general chiral VEV may be computed exactly (see e.g. 
\cite{DSEQ}--\cite{Diss}, \cite{Zah}, \cite{MSSM}),
 \beq
\langle S_{H_1}(x_1)\cdots S_{H_n}(x_n)\rangle_0 
=\Bigl( \frac{\Sigma}{2}\Bigr)^n \exp
\Bigl[ \sum_{i<j}(-)^{\sigma_i \sigma_j}4\pi D_{\mu_0} (x_i -x_j)\Bigr] 
\eeq
where $\sigma_i =\pm 1$ for $H_i =\pm$, $D_{\mu_0}$ is the massive scalar
propagator, 
\beq
D_{\mu_0}(x)=-\frac{1}{2\pi}K_0 (\mu_0 |x|), \quad \tilde
D_{\mu_0}(p)= \frac{-1}{p^2 +\mu_0^2},
\eeq
($K_0\ldots$ McDonald function) and $\Sigma$ is the fermion condensate
of the massless Schwinger model,
\beq
\Sigma =\langle\bar\Psi \Psi\rangle_0 =\frac{e^\gamma}{2\pi}\mu_0
\eeq
($\gamma\ldots$ Euler constant). The index 0 for $\mu_0$ indicates that it is
the order zero result, the index 0 for the VEVs means that they are 
computed with
respect to the massless Schwinger model. From this $Z(m,\theta)$ may be
computed (see \cite{MSSM} for details, \cite{LSm} for its physical 
implications),
\bdi
Z(m,\theta)= e^{V\alpha (m,\theta)},
\edi
\beq
\alpha (m,\theta)=\mu_0^2 \Bigl[\frac{m}{\mu_0} \frac{\Sigma}{2\mu_0}
2\cos\theta + \frac{m^2}{\mu_0^2}
\Bigl(\frac{\Sigma}{2\mu_0}\Bigr)^2 (E_+ \cos 2\theta +E_- )+
o(\frac{m^3}{\mu_0^3} ) \Bigr]
\eeq
($V\ldots$ space time volume) where $E_+$ and $E_-$ are numbers ($E_+ =-8.9139$,
 $E_- =9.7384$).

In order to compute VEVs for the massive Schwinger model one has to insert the
corresponding operators into the path integral (1), (2) and divide by the
vacuum functional $Z(m,\theta)$:
\beq
\langle \hat O \rangle_m =\frac{1}{Z(m,\theta)} \langle \hat O
\sum_{n=0}^\infty \frac{m^n}{n!}\prod_{i=1}^n \int dx_i \bar\Psi (x_i) \Psi
(x_i) \rangle_0
\eeq
Via the normalization all volume factors cancel completely, as it certainly has
to be.

It is wellknown that the Schwinger boson $\phi$ is related to the vector 
current, $J_\mu \sim \frac{1}{\sqrt{\pi}}\epsilon_{\mu\nu}\partial^\nu \phi$,
therefore, for its study vector current correlators are needed. E.g. for
the propagator $\langle\phi (y_1) \phi(y_2)\rangle_m$ and for the computation 
of the Schwinger mass one needs the $n$-point functions
\bdi
\langle\phi (y_1)\phi (y_2)\prod_{i=1}^n S_{H_i}(x_i)\rangle_0 =
\Bigl(\frac{\Sigma}{2}\Bigr)^n e^{4\pi\sum_{k<l}^n (-)^{\sigma_k \sigma_l}
D_{\mu_0}(x_k -x_l)} \cdot
\edi
\beq
\cdot  \Bigl[D_{\mu_0}(y_1 -y_2) +4\pi
(\sum_{i=1}^n (-)^{\sigma_i} D_{\mu_0}(x_i -y_2))(\sum_{j=1}^n (-)^{\sigma_j}
D_{\mu_0}(x_j -y_1))\Bigr] .
\eeq
For the Schwinger mass one finds the explicit result (\cite{SMASS})
\beq
 \frac{\mu_2^2}{\mu_0^2}=1+8\pi
\frac{m}{\mu_0}\frac{\Sigma}{2\mu_0} \cos\theta + 16\pi^2 \frac{m^2}{\mu_0^2}
\Bigl( \frac{\Sigma}{2\mu_0}\Bigr)^2 (A\cos 2\theta + B)
+o(\frac{m^3}{\mu_0^3})
\eeq
where
\bdi
A=-0.6599 \qquad ,\qquad B=1.7277.
\edi
In order to study the boson-boson interaction we need the four-point function 
$\langle \phi (y_1)\phi (y_2)\phi (y_3)\phi (y_4)\rangle_m$ and therefore the
$n$-point functions
\bdi
\langle\phi (y_1)\phi (y_2)\phi (y_3)\phi (y_4) \prod_{i=1}^n S_{H_i}(x_i)
\rangle_0 = \Bigl(\frac{\Sigma}{2}\Bigr)^n  e^{4\pi\sum_{k<l}^n (-)^{\sigma_k 
\sigma_l} D_{\mu_0}(x_k -x_l)} \cdot
\edi
\bdi
\cdot \Bigl[ D_{\mu_0}(y_1 -y_2)D_{\mu_0}(y_3 -y_4) +\mbox{ perm. } +
\edi
\bdi
4\pi D_{\mu_0}(y_1 -y_2)
\Bigl( \sum_{i=1}^n (-)^{\sigma_i}D_{\mu_0}(y_3 -x_i)\Bigr)
\Bigl( \sum_{j=1}^n (-)^{\sigma_j}D_{\mu_0}(y_4 -x_j)\Bigr)
+ \mbox{ perm. } +
\edi
\beq
16\pi^2 
\Bigl( \sum_{i=1}^n (-)^{\sigma_i}D_{\mu_0}(y_1 -x_i)\Bigr)
\Bigl( \sum_{j=1}^n (-)^{\sigma_j}D_{\mu_0}(y_2 -x_j)\Bigr)
\Bigl( \sum_{k=1}^n (-)^{\sigma_k}D_{\mu_0}(y_3 -x_k)\Bigr)
\Bigl( \sum_{l=1}^n (-)^{\sigma_l}D_{\mu_0}(y_4 -x_l)\Bigr) \Bigr]
\eeq
where "+ perm" means the sum of all {\em distinguishable} terms.
Actually the first and second types of terms are disconnected,
so only the third one will contribute.

\input psbox.tex

\section{Computation of the bound state mass}

When evaluating (7, 10) for lowest order in $m$, one finds an interaction term

$$\psboxscaled{1000}{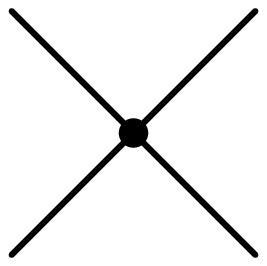}$$

\begin{center}
Fig. 1
\end{center}

\medskip

with a coupling constant (the instanton sectors $k=\pm 1$ contribute, see (1))
\beq
16\pi^2 m\Sigma \cos\theta =:2g,
\eeq
where a convenient abbreviation was introduced.

Therefore a bound state has to be expected for $|\theta |<\frac{\pi}{2}$, 
where the force is attractive, and this restriction shall be assumed in
the sequel.

Before continuing we want to fix the graphical notation:

$$\psboxscaled{800}{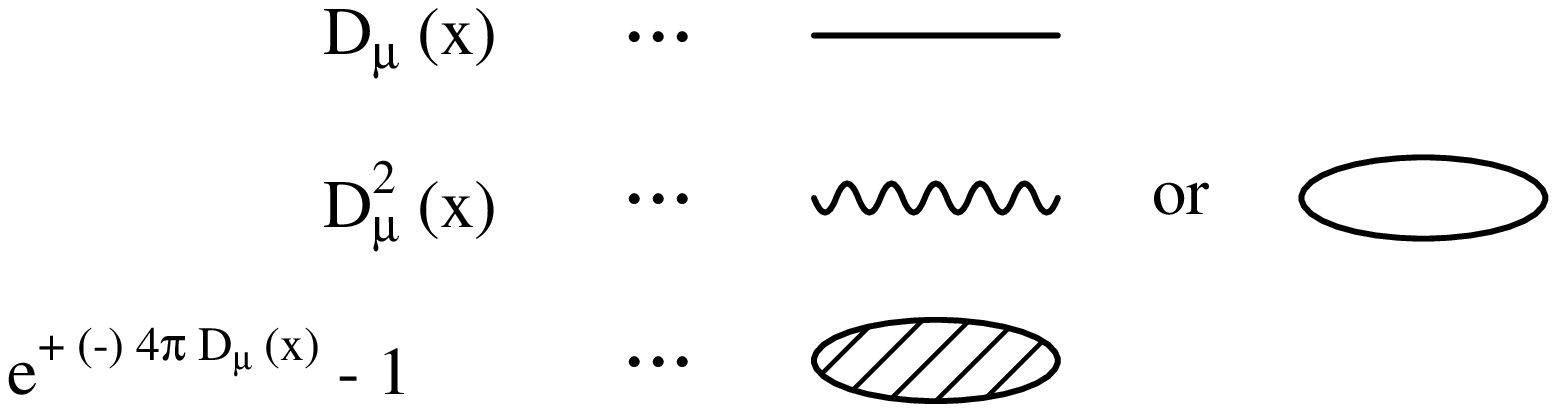}$$

\begin{center}
Fig. 2
\end{center}

\medskip

where the "$-1$" in the last expression stems from the cluster expansion
that is at the heart of the mass perturbation theory (\cite{MSSM},
\cite{SMASS}, \cite{FS1}).

Actually, we want to compute the lowest order bound state mass in an
approximation that resembles $\Phi^4$-theory, by summing a series

$$\psboxscaled{1000}{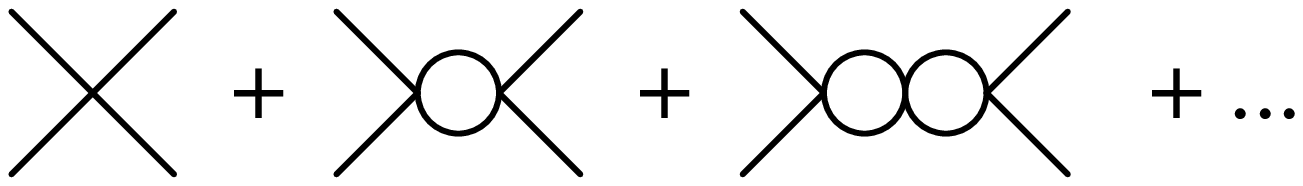}$$

\begin{center}
Fig. 3
\end{center}

\medskip

At first glance, substituting the exponential $E_{\pm}(x)=\exp (\pm 4\pi
D_\mu (x)) -1$ by the quadratic term $\frac{16\pi^2}{2!}D_\mu^2 (x)$
seems to be a very rough approximation. However, computing the bound state mass
perturbatively in $m$ will involve the solution of a self consistency equation
for a momentum (remember Euclidean conventions!) $q^2 \sim -4\mu^2
(1-\epsilon)$, where $\epsilon$ is very small for small $m$.
It is precisely the Fourier transform of $D_\mu^2 (x)$ that has a threshold 
singularity at $q^2 =-4\mu^2$, therefore it will give the main contribution
and the approximation is justified.

Next we fix $(y_1 ,y_2)$ to form the incoming state and $(y_3 ,y_4)$ to form 
the outgoing one.

Now we have to show that the graphs of the series depicted above 
(Fig. 3) actually occur
in the mass perturbation series.

First, the $\theta$ dependence is very simple. Because external lines are 
joined in pairs, and because we select the quadratic part of the
expanded exponential $\exp (\pm 4\pi D_\mu (x))$, all instanton sectors 
contribute identical terms to a given order $n$ in the mass $m$. Further the 
instanton sector $k$ gives ${n \choose n_-}$ terms, where $k=n-2n_-$.
So altogether we find
\beq 
\sum_{n_- =1}^n {n \choose n_-} e^{i(n-2n_-)\theta}=2^n \cos^n \theta
\eeq
identical terms, which shall be computed next.

In second order we have ($\mu$ is an unspecified mass for the moment)
\bdi
\frac{m^2}{2!}\Sigma^2 \cos^2 \theta \, 16\pi^2 \int dx_1 dx_2 \prod_{i=1}^4 
\Bigl[ D_\mu (y_i -x_1 )+D_\mu (y_i -x_2 )\Bigr]
\frac{(4\pi)^2}{2!}D_\mu^2 (x_1 -x_2)
\edi
\bdi
\simeq 2g\int dx_1 dx_2 \frac{m}{2!}\Sigma\cos\theta \, \frac{16\pi^2}{2}
D_\mu^2 (x_1 -x_2 )\cdot
\edi
\bdi
\cdot (D_\mu (y_1 -x_1 )D_\mu (y_2 -x_1)D_\mu (y_3 -x_2)D_\mu (y_4 -x_2)
+ (x_1 \leftrightarrow x_2 ))
\edi
\beq
= 2g\cdot g\int dx_1 dx_2 D_\mu^2 (x_1 -x_2)D_\mu (y_1 -x_1)D_\mu (y_2 -x_1)
D_\mu (y_3 -x_2)D_\mu (y_4 -x_2) 
\eeq
where the $\simeq$ indicates that the $t$ and $u$ channels have been
omitted.

For $n$-th order there are $n$ interaction points $x_i ,i=1\ldots n$.
There are $n(n-1)$ possibilities to attach the external pairs $(y_1 ,y_2)$
and $(y_3 ,y_4)$ to interaction points, and there are $(n-2)!$ 
possibilities to join these two points by a closed path of $n-1$ blobs
$\frac{16\pi^2}{2}D_\mu^2 (x_i -x_j)$. E.g. in fourth order, after
fixing an attachment of the external legs, there remain two possibilities:

$$\psboxscaled{1000}{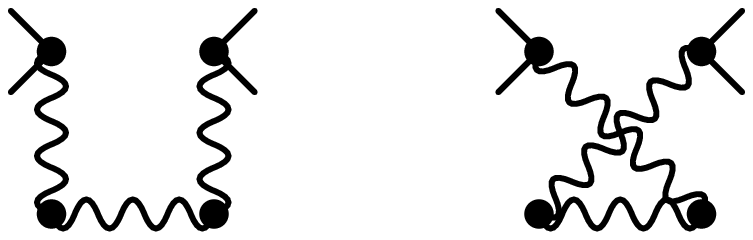}$$

\begin{center}
Fig. 4
\end{center}

\medskip

This combinatorial factor $n(n-1)((n-2)!)$ precisely cancels the $\frac{1}{n!}$
from the $n$-th order perturbation expansion. 

So all vertices have coupling constant $g$, except the first one that has $2g$ 
(actually this factor 2 may be understood, in the language of conventional
perturbation theory, as a final state symmetry factor $(y_3 ,y_4)+(y_4 ,y_3)
\ra 2(y_3 ,y_4)$).

Therefore, we indeed find a series like in Fig. 3 which, in momentum space 
and after amputation of the external legs, reads
\beq
P(q^2):=2g(1+gS(q^2)+(gS(q^2))^2 +\ldots )=\frac{2g}{1-gS(q^2)}
\eeq
where $q$ is the total incoming momentum and $S(q^2)$ is just the blob
(without external legs)

$$\psboxscaled{900}{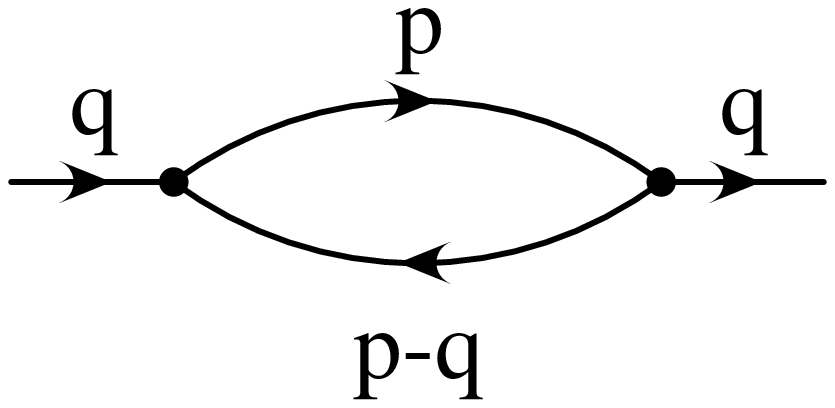}$$

\begin{center}
Fig. 5
\end{center}

\medskip

and may be evaluated by standard methods:
\bdi
S(q^2)=\int\frac{d^2 p}{(2\pi)^2}\frac{-1}{p^2 +\mu^2}\frac{-1}{(p-q)^2
+\mu^2}=\int\frac{d^2 p}{(2\pi)^2}\int_0^1 \frac{dx}{[p^2 +2pq(x-1)
+q^2 (1-x) +\mu^2 ]^2}
\edi
\bdi
=\frac{1}{4\pi}\int_0^1 \frac{dx}{q^2 x(1-x)+\mu^2} =\frac{1}{4\pi(-q^2)}
\int_0^1\frac{dy}{y^2 +(\frac{4\mu^2}{-q^2} -1)}
\edi
\beq
= \frac{1}{4\pi (-q^2)}\frac{1}{R(q^2 )}\arctan\frac{1}{R(q^2 )} ,
\eeq
\beq
R(q^2 ):=\sqrt{\frac{4\mu^2}{-q^2} -1}
\eeq
where we used the fact that, for the bound state, $-q^2$ has to be beyond 
the threshold, $-q^2 <4\mu^2$.

Now we simply have to insert this result into (14) in order to get the mass 
pole:
\beq
1=\frac{g}{4\pi (-q^2)}\frac{1}{R(q^2)}\arctan\frac{1}{R(q^2)} .
\eeq
For small fermion masses both $g$ and $R(q^2)$ are very small. Therefore,
the leading order result will stem from a matching between these two factors,
where we may set $\frac{1}{-q^2}=\frac{1}{4\mu^2}$ and $\arctan\frac{1}{R(q^2)}
=\frac{\pi}{2}$. Doing so we get
\beq
R(q^2)=\frac{g}{32\mu^2}
\eeq
or
\beq
-q^2 =4\mu^2 \frac{1}{1+(\frac{g}{32\mu^2})^2} \simeq 4\mu^2 (1-\frac{\pi^4
m^2 \Sigma^2 \cos^2 \theta}{16\mu^4})<4\mu^2 ,
\eeq
\beq
-q^2 =4\mu^2 (1-0.4892 \frac{m^2}{\mu_0^2}\cos^2 \theta +o(\frac{m^3}{\mu_0^3}))
\eeq
which is the bound state mass we are looking for. Observe that the  pole mass
equation (18) remains perfectly sensible in the limit $g\to 0$, where it leads
to the {\it exact} result $ -q^2 =4\mu_0^2 $ (no interaction).
This shows a posteriori that our approximation in Fig. 3 and formula (14)
indeed is justified for sufficiently small fermion mass.

We find from (20) that the leading bound state mass correction  
is already of second order in $m$. 
Therefore, in order to be able to express the result consistently in the
zero order Schwinger mass $\mu_0$, one should include the Schwinger mass
corrections up to second order. This corrections have been 
computed in \cite{SMASS} 
for an {\em external} boson. Because of the complicated exponential 
structure of the interaction term (see (10)) it is not obvious that the same 
corrections remain true for internal boson lines. More precisely, it is
obvious that the types of vertices are the same, it is however not completely 
obvious that even the combinatorial factors remain the same.

So let us shortly check it by investigating the perturbation formula (10).
A typical first order correction graph looks like 

$$\psboxscaled{900}{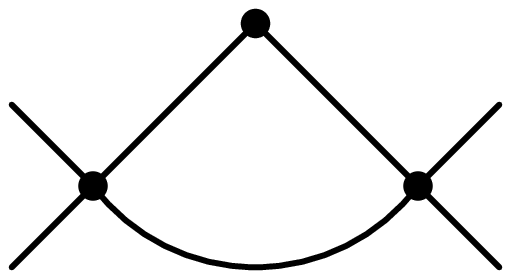}$$

\begin{center}
Fig. 6
\end{center}

\medskip

and is of third order, $\frac{m^3}{3!}$. There are $3\cdot 2=6$ possible 
attachments of the external pairs. From the exponent $\exp (\pm 4\pi D_{\mu_0}
(x_i -x_j))$ we need 3 first order coefficients instead of one second order one.
Therefore we get, in a symbolic notation (indicating integrations by
$\times $ and amputating external lines, $D\equiv D_{\mu_0}(x)$)
\beq
2g((4\pi )^3 m^2 \Sigma^2 \cos^2 \theta \, D\cdot D\times D)=
2g(g\cdot 2D(4\pi m\Sigma\cos\theta \, D\times D))=:2g(g\cdot 2D\delta_1 D).
\eeq
The second order mass correction stems from the fourth order perturbation
expansion $\frac{m^4}{4!}$. There are $4\cdot 3$ attachments, and for a 
fixed attachment each contributing type of term occurs twice. We
again need three first order propagators $4\pi D_{\mu_0}$, and one complete 
exponential $E_\pm (x)$. For a fixed attachment the following graphs
contribute:

$$\psboxscaled{800}{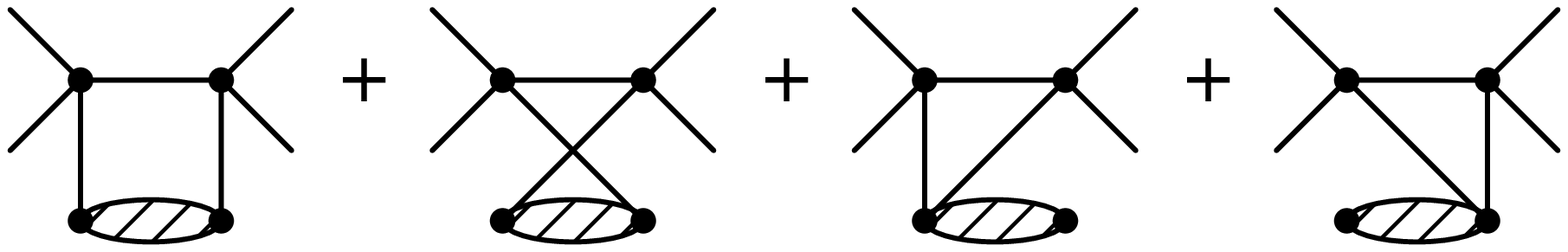}$$

\begin{center}
Fig. 7
\end{center}

\medskip

and lead to
\bdi
2gg\cdot 2D\Bigl[ 4\pi m^2 \frac{\Sigma^2}{2} \Bigl( \cos 2\theta \, 
(E_+ \cdot D\times D + E_+ \times D\times D) +E_- \cdot D\times D 
- E_- \times D\times D \Bigr) \Bigr]
\edi
\beq
=: 2g^2 \cdot 2D\delta_2 D.
\eeq
Both $\delta_1 D$ and $\delta_2 D$ are precisely like for external bosons
(see \cite{SMASS}), therefore the result (9) may be used and we find
for the bound state mass
\bdi
M_B^2 =4\mu_0^2 (1+2e^\gamma \cos\theta \, \frac{m}{\mu_0} +e^{2\gamma}
(A\cos 2\theta +B)\frac{m^2}{\mu_0^2} - \frac{\pi^2 e^{2\gamma}}{64}
\cos^2 \theta \, \frac{m^2}{\mu_0^2} +o(\frac{m^3}{\mu_0^3})
\edi
\beq
=4\mu_0^2 (1+3.5621 \cos\theta \, \frac{m}{\mu_0} +5.2361 \frac{m^2}{\mu_0^2}
-2.3379 \cos 2\theta \, \frac{m^2}{\mu_0^2} +o(\frac{m^3}{\mu_0^3})).
\eeq
This is our final result.

\section{Summary}

We have reached our aim of computing the boson-boson bound state mass 
within (fermion) mass perturbation theory. The existence of this bound state 
is a necessary consequence of the attractive force between the 
Schwinger bosons in $1+1$ dimensions.

In the actual computation we used an additional approximation besides mass 
perturbation, namely we chose the lowest possible Fock state (two bosons)
in any order of perturbation theory. This approximation could be shown
to lead to a summation of graphs like in $\Phi^4$-theory. The
approximation is justified for small fermion mass, because there this
lowest Fock state is near its threshold singularity, or, differently
stated, because it is this and only this Fock state that survives the limit
of vanishing fermion mass.

Because of the approximation the bound state pole mass remained polynomial
in the fermion mass even after the summation of all contributing graphs.
Actually it turned out quadratically in the fermion mass in leading order. 

\section*{Acknowledgement}

The author thanks Prof. Narison for the invitation to the Institute of 
Theoretical Physics of the University of Montpellier, where this work was
done, and the members of the institute for their hospitality. 

Further thanks are due to the French Foreign Ministery, the Austrian
Ministery of Research and the Austrian Service for Foreign Exchange, who
financially supported this research stay.

\end{document}